\documentclass[journal, 10 pt, twocolumn]{IEEEtran}
\IEEEoverridecommandlockouts

\usepackage{graphicx}
\usepackage{amsmath}
\usepackage{mathrsfs}
\usepackage{amsmath,xcolor}
\usepackage{array}
\usepackage{algorithm2e}
\usepackage{float}
\usepackage{subfigure}
\usepackage{cite}
\usepackage{amsmath,amssymb,amsfonts}
\usepackage{subfigure}
\usepackage{algorithmic}
\usepackage{graphicx}
\usepackage{textcomp}
\usepackage{xcolor}
\usepackage{listings}
\usepackage{color}
\usepackage{comment}
\usepackage{enumitem}
 \usepackage{graphicx}
 \usepackage{caption2}
\usepackage{subfigure}
\usepackage{multicol}

\newtheorem{theorem}{Theorem}[section]
\newtheorem{lemma}{Lemma}[section]
\def\proof{\noindent{\it Proof: }}
\def\QED{\mbox{\rule[0pt]{1.5ex}{1.5ex}}}
\def\endproof{\hspace*{\fill}~\QED\par\endtrivlist\unskip}

\newtheorem{definition}[theorem]{Definition}

\newtheorem{assumption}[theorem]{Assumption}

\newcommand{\OMIT}[1]{}

\title{\LARGE \bf
Relationship Explainable Multi-objective Optimization Via Vector Value Function Based Reinforcement Learning
}

\author{Huixin Zhan and Yongcan Cao
\thanks{The authors are with the Department of Electrical and Computer Engineering, The University of Texas at San Antonio, San Antonio, TX 78249.}
\thanks{Corresponding Author: Yongcan Cao (yongcan.cao@utsa.edu)}
}

\begin{document}

\maketitle
\thispagestyle{empty}
\pagestyle{empty}

\begin{abstract}
Solving multi-objective optimization problems is important in various applications where users are interested in obtaining optimal policies subject to multiple, yet often conflicting objectives. A typical approach to obtain optimal policies is to first construct a loss function that is based on the scalarization of individual objectives, and then find the optimal policy that minimizes the loss. However, optimizing the scalarized (and weighted) loss does not necessarily provide guarantee of high performance on each possibly conflicting objective. In this paper, we propose a vector value based reinforcement learning approach that seeks to explicitly learn the inter-objective relationship and optimize multiple objectives based on the learned relationship. In particular, the proposed method is to first define correlation matrix, a mathematical representation of the inter-objective relationship, and then create one actor and multiple critics that can co-learn the correlation matrix and action selection. The proposed approach can quantify the inter-objective relationship via reinforcement learning when the impact of one objective on another is unknown \textit{a prior}. We also provide rigorous convergence analysis of the proposed approach and present quantitative evaluation of the approach based on two testing scenarios. 
\end{abstract}

\section{Introduction}
In recent years, the application of reinforcement learning in missions with high-dimensional sensory inputs has shown the potential of creating artificial agents that can learn to accomplish a number of challenging tasks, including the Atari games~\cite{mnih2015human,guo2014deep,schaul2015prioritized,wang2016dueling,van2016deep,oh2015action,nair2015massively}, self-driving cars~\cite{pan2017virtual}, and Go~\cite{maddison2014move,silver2016mastering,silver2018general}. The approaches developed therein mainly focus on the settings when a single objective needs to be optimized. In reinforcement learning, most research seeks to find a single usable strategy, without considering the trade-off among potential alternatives that may increase one objective's value at the cost of another.

In the presence of multiple objectives, reinforcement learning algorithms have been proposed to address the case when various objectives may need to be optimized. This type of research has gained more interest recently~\cite{vamplew2011empirical,roijers2013survey,tesauro2008managing}. In practical applications, the completion of a mission requires the simultaneous satisfaction of multiple objectives such as balancing the power consumption and performance in Web servers~\cite{tesauro2008managing}, and driving the growth in multi-objective research~\cite{vamplew2011empirical,roijers2013survey}. Such problems can be modeled as multi-objective Markov decision processes (MOMDPs), and solved by some existing multi-objective reinforcement learning (MORL). However, solutions obtained via these approaches can hardly balance the possibly conflicting objectives to achieve satisfactory performance on all objectives.

Several interesting MORL approaches that have been developed include: (1) a multi-objective deep RL framework, (2) a modular multi-objective deep RL (MODRL) with decision values, and (3) Softmax exploration strategies for MORL. The multi-objective deep RL framework proposed the use of the linear weighted sum and the nonlinear thresholded lexicographic ordering methods to develop their MODRL framework that includes both single and multi-policy strategies~\cite{nguyen2018multi}. The modular MODRL with decision values proposed an architecture in which separated deep Q-networks (DQNs) are used to control the agent's behavior with respect to particular objectives. Each DQN has an additional decision value output that acts as a dynamic weight used while summing up Q-values from particular DQNs~\cite{tajmajer2018modular}. The Softmax exploration strategies for MORL used softmax-epsilon selection based on a non-linear action-selection operator. In this process, the agents incorporate an action-selection function that is defined as an ordering over these Q-values~\cite{vamplew2017softmax}. In summary, most of the algorithms are based on the scalarization method to transform the multi-objective problem into a single objective one. The scalarization can be nonlinear or linear~\cite{van2013scalarized}. Other advanced methods include the convex hull and varying parameters approaches~\cite{liu2015multiobjective}.

A popular scalarization method is the deep optimistic linear support learning algorithm (DOL). DOL computes the weights that are used to generate a scalarized objective using convex coverage set~\cite{roijers2015computing}. Once the weights are obtained, the scalarized objective is calculated as the inner product between the weight vector and the objective vector. However, the standard DOL does not address the (potential) conflicting nature among objectives directly and still uses a scalar value function.

\begin{figure}
\begin{multicols}{2}
\centering
\subfigure[$t_{s}$ = 25]{\includegraphics[width=3.5cm,height=1.9cm]{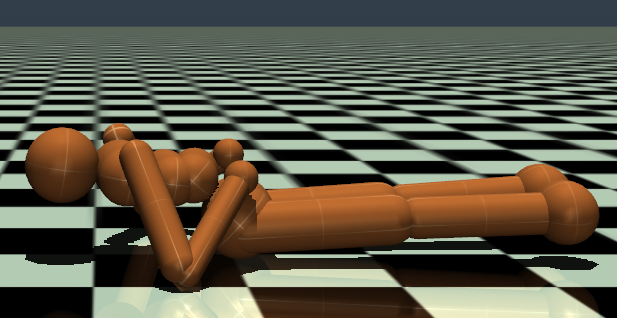}}
\subfigure[$t_{s}$ = 90]{\includegraphics[width=2.0cm,height=1.9cm]{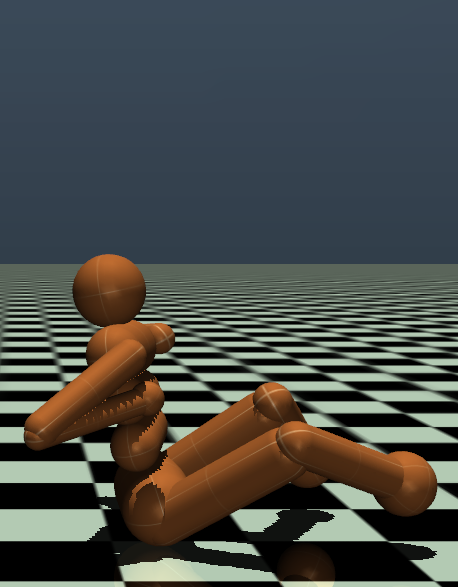}}\\
\subfigure[$t_{s}$= 145]{\includegraphics[width=1.8cm,height=1.9cm]{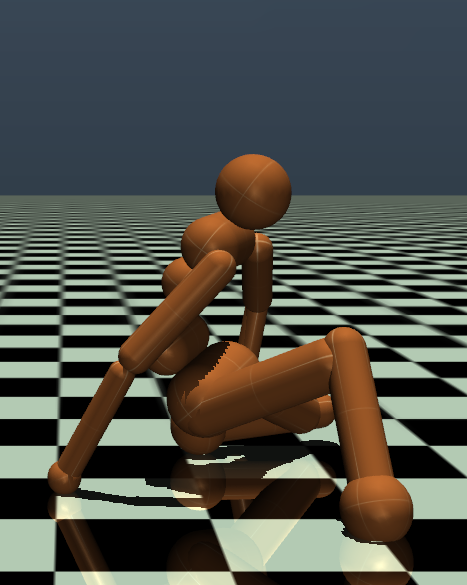}}
\subfigure[$t_{s}$ = 165]{\includegraphics[width=1.8cm,height=1.9cm]{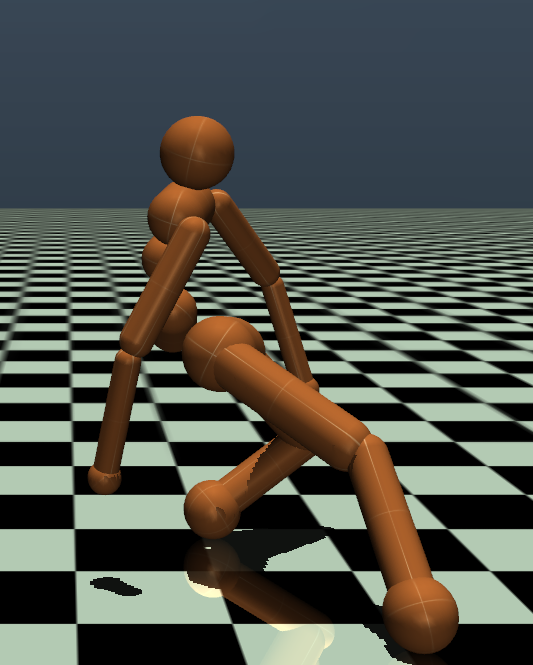}}
\subfigure[$t_{s}$ = 210]{\includegraphics[width=1.8cm,height=1.9cm]{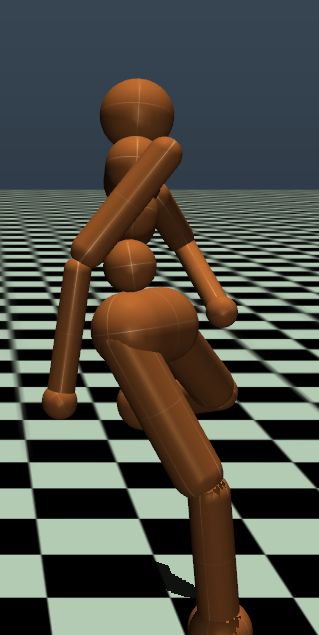}}

\hspace{-.4in}
\subfigure[$t_{s}$ = 3635]{\includegraphics[width=2.7cm,height=4.5cm]{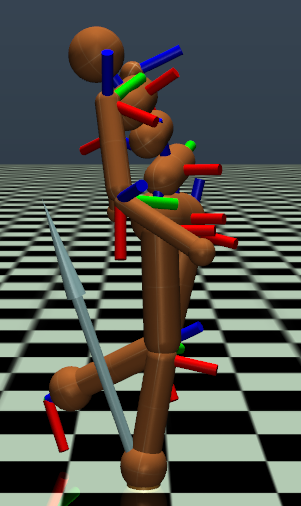}}
\subfigure[$t_{s}$ = 3710]{\includegraphics[width=2.7cm,height=4.5cm]{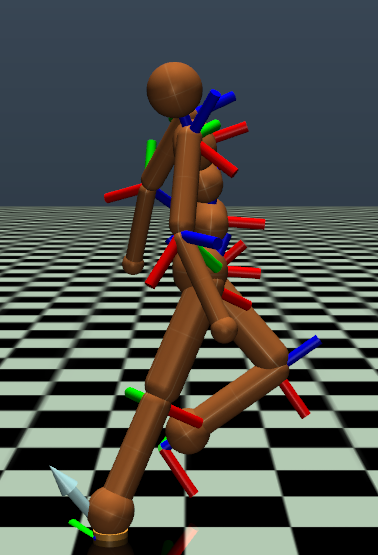}}
\end{multicols}
\caption{Sample frames from an experiment based on the policy obtained from the proposed method in a MuJoCo environment. $t_{s}$ represents the time step.}
\label{MuJoCo}
\end{figure}

When the objective state values for all objectives are considered as a vector and balancing them is required, action selection can become very challenging. To address this issue, this paper focuses on proposing a new vector value function based deep reinforcement learning method. The proposed research has three main contributions. \underline{First}, instead of using scalarized Q-value and the action selection approach based on the priority objective value, the proposed method supports vectorized objective state values. Then all objective state values are used to train the critics sequentially while the action selection is based on an actor network. \underline{Second}, by explicitly defining the inter-objective relationship via a parametric form, each objective state value can be computed based on its own value and other objective state values. To identify the relationship among objectives, we propose a new concept, called marginal weight, and use it to compute the parameters in the parametric form. The obtained weights, quantifying the inter-objective relationship, are then used in the training of actor critic to update the action policy. \underline{Third}, our method is applicable in high-dimensional continuous action spaces. As an example, Fig.~\ref{MuJoCo} shows the testing results when using the proposed method in the MuJoCo environment. To our best knowledge, this is the first time that actor critic with quantifiable inter-objective relationship is developed to solve MORL using vector value functions. We also show via two examples that the proposed method outperforms the existing single objective optimization methods.

The remainder of the paper is organized as follows. Section~\ref{sec:problemformulation} provides a brief problem formulation. Section III describes the main approach. Section IV describes the testing results. Finally, section V provides a brief summary with discussion on potential future directions.

\section{Problem Statement}\label{sec:problemformulation}

%

In this paper, we consider the problem when multiple objectives $O_i,~i=1,\cdots,I,$ need to be optimized for a given mission, where $I$ denotes the number of objectives. For example, in robotic locomotion, maximizing forward velocity but minimizing joint torque and impact with the ground, result in a very large number of options to consider. Let a control policy that generates the actual values for these objectives be given by $\pi$. We use $V_i^\pi, i=1,\cdots, I,$ to represent the value functions for $O_i,~i=1,\cdots,I,$ subject to the control policy $\pi$. A typical approach to optimize objectives $O_i,~i=1,\cdots,I,$ is to construct a scalarized value function of the form
\begin{equation}\label{eq:scalar_function}
V_{w}^{\pi}=\sum_{i=1}^Iw_iV^\pi_i,
\end{equation}
where $w=[w_1,\cdots, w_I]$ satisfying $w\textbf{1}=1$, where $\textbf{1}$ is an all-one column vector, and $w_i$ specifies how much each objective contributes to the scalarized objective. A more general form of the scalarized value function is given by $V_{w}^{\pi}=f\left(w, V^{\pi}\right)$ ~\cite{roijers2013survey}, where $f(\cdot,\cdot)$ is a nonlinear function. Hence, a multi-objective optimization problem can be converted to a single-objective optimization problem. To derive the optimal policy $\pi$ for the single-objective optimization problem under unknown environments, which are typically modeled by Markov Decision Processes (MDP), one important method is to formulate the problem in the form of Bellman equation as:
\begin{equation}
V_w^{\pi}\left ( s \right )=\sum_{a\in \mathcal{A}} \pi\left ( a|s \right )\left ( \mathcal{R}_{s}^{a} + \gamma \sum_{s^{'}\in S} \mathcal{P}_{ss^{'}}^{a}V_w^{\pi}\left ( s^{'} \right ) \right ),
\end{equation}
where $\pi(a|s)$ is the probability that an action $a$ is taken given current state $s$ subject to policy $\pi$, $\mathcal{A}$ is the action space, $\mathcal{R}_s^a$ is the immediate reward of $V_\omega$ when taking action $a$ at state $s$, $\gamma$ is the discounting factor, $\mathcal{P}_{ss^{'}}^{a}$ is the state transition probability from $s$ to $s'$, and $V_w^{\pi}\left ( s^{'} \right )$ is the expected reward under policy $\pi$ with initial state $s'$. 
In the typical actor-critic method, we iteratively run the policy to generate samples, fit a model to estimate the value function, and then improve the policy by policy gradient to obtain the optimal policy.

Although the obtained optimal policy $\pi$ may yield the largest value for the weighted sum of all objectives, it may not optimize any individual objective because at most one objective is directly optimized when $w_i=1$ for some $i$ and $w_j=0,~\forall j\neq i$. In addition, the relationship among these objectives is not explicitly identified such that the weights can be selected more optimally. To address the challenge, our objective here is to provide more direct and precise optimization of multiple objectives simultaneously while explicitly quantifying the relationship among these objectives.

\section{Main Approach}\label{sec:MP}

For a policy $\pi$, we here propose the construction of a set of \textit{multi-objective vector value functions} $\mathcal{Y}^\pi=[\mathcal{Y}_{1}^{\pi}, \cdots, \mathcal{Y}_{I}^{\pi}]\in\mathbb{R}^I$ , whose $i$th entry $\mathcal{Y}_{i}^\pi$ is the expected weighted sum of all objective state values $V_j^\pi,~j=1,\cdots,I,$ characterizing one specific objective state value and how other objectives impact this objective under policy $\pi$. 
Let $y_{m}^{i}$ be a specific value of $\mathcal{Y}_{i}^{\pi}$. To solve Bellman equation and find a policy that optimizes the objective value, we make the following two assumptions.

\begin{assumption}
Objective state values $V_{1}^{\pi}, \cdots, V_{n}^{\pi}$ have additive dependent property, \textit{i.e.}, the objective values have an additive form.
\end{assumption}
\begin{assumption}
Each multi-objective value function $\mathcal{Y}_{i}^{\pi}$ has a linear form.
\end{assumption}

Given these assumptions, we will present a formal definition that quantifies how each objective state value $V_i^\pi$ acts on multi-objective vector value function $\mathcal{Y}_{i}^{\pi}$ under policy $\pi$.

\begin{definition}\label{definition2}
Each multi-objective value function is a cumulative sum of objective state values with additive specific impact elements:
\begin{equation}\label{eq:yipi}
{\mathcal{Y}_{i}^{\pi}\left[ik\right]}
=\sum_{j=1}^I w_{ij}[ik]V_j^\pi[ik],\quad i=1,\cdots,I,
\end{equation}
where $w_{ij}$ is the weight quantifying the impact of $V^\pi_j$ on $V^\pi_i$, $ik$ is the time step number, and $k$ is the number of sequence.
\end{definition}

In Definition \ref{definition2}, a scalar $\mathcal{Y}_{i}^{\pi}$ is an inner product of the objective state value vector $V_{\cdot}^{\pi}$ and the basis vector $w_{i\cdot}$. 
Define 
\begin{equation}\label{eq:CM}
W= \left [ w_{ij} \right ]\in\mathbb{R}^{I\times I}.
\end{equation} 
Matrix $W$ is called \textit{correlation matrix}.

Because the impact of one objective on another objective is unknown \textit{a priori}, the correlation matrix $W$ needs to be updated based on the input observation, denoted as $\mathcal{X}$, and a collection of objective value spaces, denoted as $\left \{ \mathcal{Y}^{i} \right \}_{i\in \left | I \right |}$, for each objective value $y^{i}$ with $m$ input/output examples denoted as $\left(x^{i}_{1},y^{i}_{1}\right),....,\left(x^{i}_{m},y^{i}_{m}\right)$. Correlation matrix is updated via numerous batches. During each batch,  $I\times k$ time steps will be divided into $k$ sequences. In the $ik$th mini-batch, $m$ examples are trained to fit $y^i$, which is then used to update the objective state values $V_i$. $\left\{V_j^\pi\left[ik\right]\right\}_{j\in \left | I \right |}$ is then used to update the $i$th row of the correlation matrix $W$. Next we will explain how the udpate procedure works in detail.

Before discussing the detailed procedure, we will first introduce a few definitions that are needed in solving multi-objective optimization problems using (deep) reinforcement learning. Let a trajectory $\tau^{i}=\left\{q\left(x_{1}\right),q\left(x_{t+1}|x_{t},a_{t}\right),H\right\}$ consist of a distribution over initial observations $q\left(x_{1}\right)$ with a transition distribution $q\left(x_{t+1}|x_{t},a_{t}\right)$ and an episode length $H$. We define the loss $\mathcal{L}\left(x_{1},a_{1},...,x_{H},a_{H}\right)$ as the negated expected accumulated reward for a series of state-action pairs with length $H$ given by
\begin{equation}\label{eq:loss}
\mathcal{L}^{i}\left(f_{\theta_{\pi}}\right)=-E_{x_{t},a_{t}\sim f_{\theta_{\pi},\tau^{i}}}\left[\sum_{t=1}^{H}R_{i}\left(x_{t},a_{t}\right)\right],\quad i=1,\cdots,I
\end{equation}
where $E$ is the expectation operation, $f_{\theta_{\pi},\tau_{i}}$ is the action distribution function determined by the policy $\pi$ that is assumed to be constructed using neural network with $\theta_{\pi}$ acting as the hyperparameter. Another set of hyperparameter is needed to quantify the map from $V_{i}^{\pi}$ to $\mathcal{Y}^{i}$ defined as
\begin{equation}\label{eq:hyper_V_Y}
f\left(V_{i}^{\pi};w_{ii},w_{ij}\right): V_{i}^{\pi} \rightarrow \mathcal{Y}^{i},
\end{equation}
where $w_{ii}$ and $w_{ij}$ are the weights in the $i$th row of the correlation matrix $W$.

\subsection{$W$ Update}\label{subsec:wab}


We now describe how $W$ is updated within a batch. First, let\rq{}s define the map from $\mathcal{X} $ to $\left \{ \mathcal{Y}^{i} \right \}_{i\in \left | I \right |}$ in a parametric form as $f^{i}\left(x;\theta_{V_{i}},\theta_{V_{j}}\right) : \mathcal{X} \rightarrow \left \{ \mathcal{Y}^{i} \right \}_{i\in \left | I \right |}$, where $\theta_{V_i}$ and $\theta_{V_j}$ are the aggregated hyperparameters of $\theta_{\pi}$ and $w_{ij}$. The main idea to update $W$ is to first obtain the undominated set $US$, then evaluate the marginal weights 
on the undominated set, and finally use the best marginal weight to udpate $W$.

Before discussing how to obtain undominated set, let\rq{}s first explain what is the undominated set and how to obtain it. A formal definition of the undominated set is given in the following definition:
\begin{definition}
The undominated set, denoted as $US$, is the set of all actions and associated payoff values that are optimal for some $w$ of the scalarization function $f\left(w, V^{\pi}\right)$\cite{desideri2012mutiple}
\begin{equation*}
U\left (a,V \right )=\left \{ (a,u_w(a))|\exists w~\textup{such that}~u_{w}\left ( a \right )\geq u_{w}\left ( a^\prime \right ),\forall a^\prime\right \},
\end{equation*}
where $u_w(\cdot)$ is the value when taking action $a$ based on the weight $w$.
\end{definition}

The introduction of undominated set is to quantify the relationship between different policies subject to a number of objectives because one policy can yield good performance for one objective while poor performance for another objective. For instance, for objective $i_{1}$, solution $\theta$ is better when $\mathcal{\hat{L}}^{i_{1}}\left ( \theta_{V_{i_{1}}}, \theta_{V_{j_{1}}}\right )< \mathcal{\hat{L}}^{i_{1}}\left ( \overline{\theta}_{V_{i_{1}}}, \overline{\theta}_{V_{j_{1}}}\right )$, while for objective $i_{2}$, solution $\overline{\theta}$ is better when $\mathcal{\hat{L}}^{i_{2}}\left ( \theta_{V_{i_{2}}}, \theta_{V_{j_{2}}}\right )>  \mathcal{\hat{L}}^{i_{2}}\left ( \overline{\theta}_{V_{i_{2}}}, \overline{\theta}_{V_{j_{2}}}\right )$. $US$ defines the set of vector objective state values that the optimal value must reside in because every other vector objective state value not in the set will not be the best choice since there exists at least one vector in $US$ that is not smaller than it. After we define the undominated set, we introduce the definition of marginal weight.

\begin{definition}\label{mw}
Marginal weight (MW), a set of weights derived from the undominated set $U(a,V)$, is defined as
$
MW(V)\in \left \{ w| \tilde{a}\in U(a,V), u_{w}\left ( \tilde{a} \right )\geq u_{w}\left ( a^\prime \right ),\forall a^\prime \right \}. 
$
\end{definition}

Because it is difficult to obtain the undominated set directly, we here employ the approximate optimistic linear support (AOLS) approach \cite{roijers2014bounded} to get an approximated undominated set. We next describe AOLS in mode detail.


The AOLS is a method that can gradually improve the approximation of the undominated set (US). Given a maximum improvement threshold $\varepsilon>0$, the AOLS algorithm can compute an approximated $\varepsilon$-optimal undominated set, denoted as $\overline{US}$, which may diverge from the optimal undominated set by at most $\varepsilon$. Consequently, its \textit{marginal weight} can be obtained. 
Before a complete undominated set is obtained, a partial undominated set can be obtained by evaluating the largest improvement for weights via the priority queue of the marginal weight in this step.
An element in the vector value function over a partial US is defined by 
$
V_{S}^{*}\left ( w \right )=\max_{V^{\pi}\in S}\, w\cdot V(s,\phi_V),
$
where $S$ is the partial undominated set, $V(s,\phi_V)$ is the approximated objective state value vector based on the current critic networks using the current hyperparameter $\phi_V$, and $s$ is the current state.


\begin{figure}[htbp]
\begin{center}
\includegraphics[width=7.5cm]{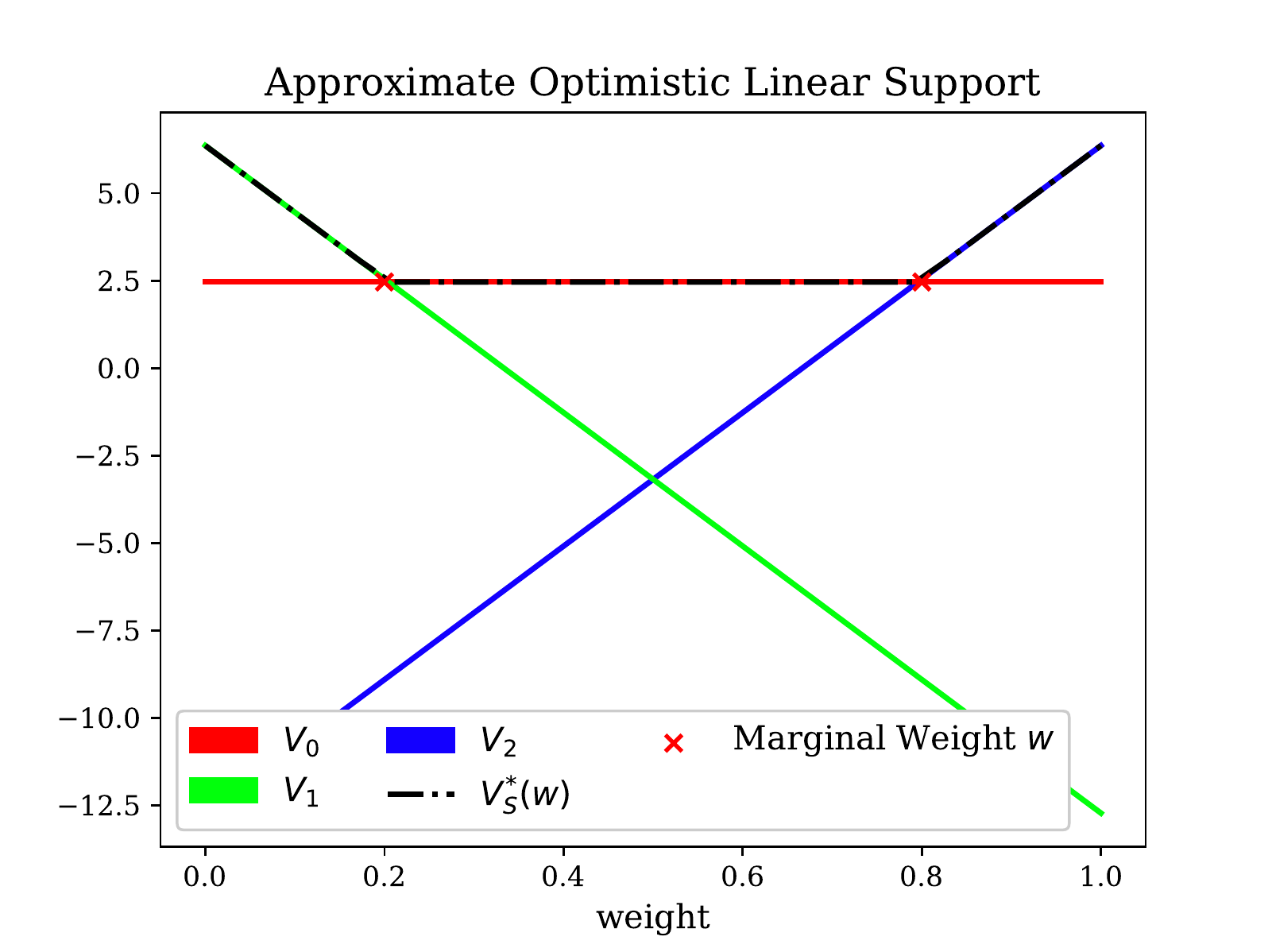}
\caption{An example of $V_{S}^{*}\left ( w \right )$ for a $S$ with two objectives.}
\label{aolsf}
\end{center}
\end{figure}

In the presence of two objectives, an example of $V_{S}^{*}\left ( w \right )$ for a $S$ containing three value vectors is shown in Fig.~\ref{aolsf}. $V_{S}^{*}\left ( w \right )$ is a piecewise linear and convex function that consists of line segments, each of which is the upper surface among all scalarized value functions. The marginal weight is the set of weights corresponding to the corners of the convex upper surface marked with red `x'. In the presence of three objectives, each of the element in $V_{S}^{*}\left ( w \right )$ associated with a policy is a plane instead of a line. When there are more than three objectives,  each element in $V_{S}^{*}\left ( w \right )$ can be represented as a hyperplane. AOLS always selects the marginal weight $w$ that maximizes an optimistic upper bound on the difference between $V_{\overline{US}}\left ( w \right )$ and $ V_{S}^{*}\left ( w \right )$, \textit{i.e.}, $V_{\overline{US}}\left (w \right )- V_{S}^{*}\left ( w \right )$, which can be updated iteratively to obtain a more accurate $\max_{w}  V_{US}(w)$. The pseudocode for AOLS is shown in the Appendix A.1. 

The previous part of this subsection describes how one row of $W$ is updated in a sequence. In order to update all rows of $W$, $I$ sequences are needed. To do so, one batch with $k$ time steps is divided into $I$ sequences. After updating all $I$ rows of $W$ using one batch with $I$ sequences, the entire $W$ will be updated. As a consequence, the vector value function $\mathcal{Y}^{\pi}=[\mathcal{Y}_{i}^{\pi}]\in\mathbb{R}^{I\times I}$, where $\mathcal{Y}_{i}^{\pi}$ is defined in \eqref{eq:yipi}, can be updated via
\begin{equation}\label{eq:movf}
\begin{aligned}
\begin{bmatrix}
\mathcal{Y}_{1}^{\pi}\left[k\right]
\\
\vdots
\\
{\mathcal{Y}_{n}^{\pi}\left[k\right]}
\end{bmatrix}
=\begin{bmatrix}
w_{11}\left[k\right] & w_{12}\left[k\right] & \cdots  &w_{1n}\left[k\right] \\
\vdots  & \vdots  & \ddots  & \vdots \\
w_{n1}\left[k\right] &w_{n2}\left[k\right]  &\cdots   & w_{nn}\left[k\right]
\end{bmatrix} \begin{bmatrix}
V_{1}^{\pi}\left[k\right] 
\\
\vdots
\\
{V_{n}^{\pi}\left[k\right] }
\end{bmatrix}.
\end{aligned}
\end{equation}


\subsection{$V$, $\mathcal{Y}$, and $\pi$ Update}\label{moac} 

After obtaining $W$ based on the procedure described in Subsection~\ref{subsec:wab}, we will describe how to update $V$, $\mathcal{Y}$, and $\pi$. We first use $W$ to update $\mathcal{Y}_i^\pi$, which is then used to train a value function approximation network to obtain policy network for an updated policy $\pi$. The newly obtained actor network is then used to generate new $V_i^\pi$ for the continuous update of $W$ described in Subsection~\ref{subsec:wab}. The following of this subsection will provide more detailed description.

Based on \eqref{eq:movf}, the new value of $\mathcal{Y}^{\pi}$ can be obtained. To obtain policy network and value function approximation network, we propose to adopt an actor-critic network with \textit{one actor network} and \textit{$I$ critic networks}, where the actor network is used to maximize the objective state value and each critic network is used to map from the state action pair to $\mathcal{Y}_{i}^{\pi}$. Assume that the actor network with hyperparameter $\theta_\pi$ generates action via
$a = \pi \left(s; \theta_{\pi}\right).$
The hyperparameter $\theta_{\pi}$ can be updated using policy gradient given by~\cite{vamvoudakis2010online}:
\begin{align*}
\Delta \theta_{\pi}&\sim\sum_{k}\bigtriangledown_{\theta _{\pi}}\log \pi_{\theta}\left ( s_{k},a_{k} \right ) \delta_{k,t},
\end{align*}
where $\delta_{k,t}$ is the expected value of the $i$th objective, also known as the temporal difference (TD) residual of $\widehat{V}^\pi_{i}$ with discount $\gamma$~\cite{sutton2018reinforcement}, given by
\begin{align}\label{eq:aa}
\delta_{k,t} = &r_{i}\left ( s_{k,t},a_{k,t} \right )+\gamma\widehat{V}_{i}^{\pi(\theta_\pi)}\left ( s_{k,t+1};\phi_{V_{i}} \right )-\widehat{V}_{i}^{\pi \left( \theta^{-}_{\pi} \right)}\left ( s_{k};\phi^{-}_{V_{i}} \right )
\end{align}
where $r_{i}\left ( s_{k,t},a_{k,t} \right )$ is the immediate reward at the $t$th time step on the $k$th experience, $\widehat{V}_{i}^{\pi \left( \theta^{-}_{\pi} \right)}\left ( s_{k,t};\phi^{-}_{V_{i}} \right )$ is the approximation of the value function $V_i$ based on the old hyperparameter $\theta^{-}_{\pi}$ for the actor network and the old hyperparameter $\phi^{-}_{V_{i}}$ for the $i$th critic network, and $\widehat{V}_{i}^{\pi(\theta_\pi)}\left ( s_{k+1};\phi_{V_{i}} \right )$ is the approximation of the value function $V_i$ based on the updated hyperparameter $\theta_{\pi}$ for the actor network and the updated hyperparameter $\phi_{V_{i}}$ for the $i$th critic network.


For the $I$ critic networks, its $i$th neural network with hyperparameter $\phi_{V_{i}}$ is used to approximate each element in the vector value function  $V_{i}^{\pi}\left(s\right)$. Assume that the critic function is given by $V_{i}\left(s;\phi_{V_{i}}\right)$ with $\phi_{V_{i}}$ serving as the hyperparameter. The hyperparameter can be updated via
$
\Delta _{k}\phi _{V_{i}} \sim -\triangledown_{\phi _{V_{i}}} \sum_{k} \delta^2_{k,t}.
$

In the standard TD-residual method, the value of one action evaluated via \eqref{eq:aa} is an incremental form of value iteration. The key drawback of the standard TD-residual method includes the need for a large number of samples and large variance of policy gradient estimate. To address these issues, an existing approach, called generalized advantage estimator (GAE)~\cite{schulman2015high}, can be used to evaluate the action advantages and perform the policy updates using proximal policy optimization~\cite{schulman2017proximal,schulman2015trust,rockafellar1991scenarios}. The GAE is defined by
\begin{align*}
\widehat{A}_{t}^{GAE\left(\gamma ,\lambda \right)} =&\lim\limits_{H\to\infty} (1-\lambda)\sum_{j=1}^H \widehat{A}_{t}^{\left(j\right)} \\=&\lim\limits_{H\to\infty} ( 1-\lambda ) \sum_{j=1}^H \lambda^{j-1} \sum_{k=1}^j \gamma^{j-1}\delta_{k,t+j-1}\\ 
= &\sum_{l=0}^{H}\left(\gamma \lambda \right)^{l}\delta_{k,t+l},
\end{align*}
where $\lambda\in \left[0,1\right]$ and $\gamma\in \left[0,1\right]$ adjusts the bias-variance tradeoff of GAE. 
 
After new hyperparameters of the advantage actor-critic network models are obtained, $V^{\pi}$ can be obtained via new samples using the updated policy. Afterwards, the procedure in Subsection~\ref{subsec:wab} can be implemented to obtain the updated $W$. The entire process will iterate until $V_{\overline{US}}\left ( w \right )- V_{S}^{*}\left ( w \right )<\epsilon$. The pseudocode for the proposed algorithm is given in the Appendix A.2. The convergence property of the proposed policy is described in the following theorem. 

\begin{theorem}\label{th:main}
The proposed policy will converge with probability $1$ when the actor updates much slower that the critic.
\end{theorem}
\proof See Appendix A.3. \endproof 

\section{Experimental Evaluation}

In this section, we evaluate the performance of the proposed vector value function based multi-objective optimization method. We first describe the experiment setup, then demonstrate how the relationshiop among various objectives can be quantified via a correlation matrix. After that, we will show how the maximal relative improvement 
$\Delta_{r}\left(w\right)=\frac{V_{\overline{US}}(w) - V_{S}^{*}(w)}{V_{\overline{US}}\left(w\right)}$ 
changes with respect to the number of training episodes. Finally, we show the the testing results and the solution stability. 
\subsection{Setup}

We here use two testing environments: (1) Humanoid-v2, and (2) HumannoidStandup-v2, on the MuJoCo physics engine~\cite{todorov2012mujoco}. For Humanoid-v2, we select five objectives: Mean Episode Length (MEL), Mean Episode Reward (MER), Linear Velocity (LVel), Quadratic Control (QCtrl), and Quadratic Impact Cost (QIm). For HumannoidStandup-v2, we select three objectives: Standup Cost, Quadratic Control, and Quadratic Impact Cost.

In both scenarios, we adopt a neural network policy with two hidden layers of size 64 using ReLU as the activation function. We use the proximal policy optimization clipping algorithm with $\epsilon=0.2$ as the optimizer. The critics have one hidden layer of size 64. The discounting factor is selected as $\gamma=0.99$. One episode, characterizing the number of time steps of the vectorized environment per update, is chosen as $2048$. For stabilization purposes, we execute parallel episodes in one batch. The batch size is chosen as the product of the episode size and the number of environment copies simulated in parallel. For Humanoid-v2, he number of environment copies is selected as $8$. For  HumannoidStandup-v2, the number of environment copies is selected as $2$. The parameters are optimized using the Adam algorithm~\cite{kingma2014adam} and a learning rate of $3\times10^{-4}$. All of the experiments were performed using TensorFlow, which allows for automatic differentiation through the gradient updates~\cite{abadi2016tensorflow}. 

\subsection{Correlation Matrix}\label{CM}
As stated in Subsection~\ref{subsec:wab}, $W$ can be estimated dynamically by computing the marginal weight via $I$ sequences, where $I$ is the number of objectives. For example, for Humanoid-v2, we consider $5$ objectives, \textit{i.e.,} $I=5$. We separate the batch size, \textit{i.e.,} the number of steps of the vectorized environment per update, into five sequences equally. These five sequences correspond to five objectives: MEL, MER, LVel, QCtrl, and QIm. In each sequence, only the $i$th multi-objective value $\mathcal{Y}_{i}^{\pi}\left[ik\right]$ is fitted via regression based on the mean-squared error. The parametric hypothesis per objective is considered in the form of $f^{i}\left(x;\theta_{V_{i}},\theta_{V_{j}}\right) : \mathcal{X} \rightarrow \left \{ \mathcal{Y}^{i} \right \}_{i\in \left | I \right |}$, in which $\theta_{V_{i}}$ is the self dependency parameter while $\theta_{V_{j}}$ is the cross-objective dependency parameter defined in \eqref{eq:hyper_V_Y}. In other words, $w_{ii}$ characterizes the self dependency of objective $i$ while $w_{ij}$ characterizes the impact of objective $i$ on objective $j$. By arranging $w_{ij}$ according to \eqref{eq:CM}, we can obtain a correlation matrix at each sequence with size $I\times I$. For Humanoid-v2, the correlation matrix is a $5\times 5$ matrix.


One example of the correlation matrix after $425$ sequences is given by
\begin{center}
$W=
\begin{bmatrix}
1. & 0. & 0. & 0. & 0. \\
0. & 1. & 0. & 0. & 0. \\
0. & 0. & 0.705 & 0.208 & 0.087 \\
0. & 0. & 0.184 & 0.754 & 0.061\\
0. & 0. & 0.090 & 0.013 & 0.897
\end{bmatrix}.$
\end{center}
From $W$, we can observe that the first two objectives MEL and MER, corresponding to the first two rows in W, are independent. In other words, MEL and MER will be not affected by other objectives. It can also be observed that the last three objectives, namely QCtrl, LVel and QIm, are dependent. In particular, the third row of $W$ indicates that objective $3$ has direct impact on objectives $4$ and $5$. Similarly, the fourth and fifth rows of $W$ indicate that objectives $4$ and $5$ have direct impact on other objectives except the first two objectives.
In other words, QCtrl, LVel, and QIm are conflicting. Moreover, QCtrl has larger impact on LVel than that on QIm because $w_{34}=0.208>0.087=w_{35}$. Hence, the relationship among these objectives is explicitly described by the correlation matrix. The multi-objective value function taking a vector-valued form reflects the weighted sum of all objectives based on the impact of each objective on other (possibly conflicting) objectives. Each element of the correlation matrix will gradually converge to a value with very small variance.

\subsection{Accuracy vs Episodes}\label{AE}

\begin{figure}[htbp]
\begin{center}
\includegraphics[width=7cm]{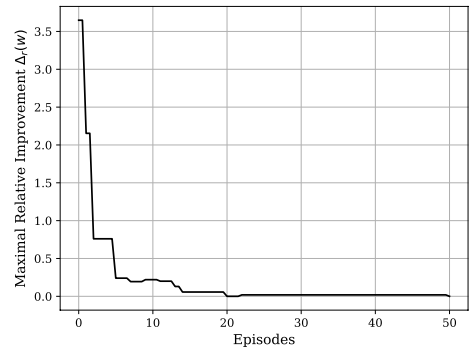}
\caption{Evolution of $\Delta_{r}\left(w\right)$ with respect to episodes in percentage.}
\label{fig:Evol}
\end{center}
\end{figure}

We further investigated the effects of the number of training episodes (including series of time steps) on the maximal relative improvement of the undominated set, defined at the beginning of the section. Fig.~\ref{fig:Evol} shows how the maximal relative improvement of the undominated set evolves with respect to the number of episodes. It can be seen from Fig.~\ref{fig:Evol} that the error is highly affected by the number of training episodes. Although the proposed method is unable to provide sufficient accuracy to build the undominated set initially, the deviation will gradually decrease to $0$ when the number of episodes is $20$. 

\subsection{Testing Results and Discussion}\label{Test}
We now present the main testing results based on the proposed method presented in Section~\ref{sec:MP}. For simplicity of presentation, we call our proposed method the PPO\_MW method. To show the benefit of the proposed PPO\_MW method, we will also show the results when (1) multi-objective optimization is solved via one single-objective optimization, and (2) the marginal weight in our method is replaced by the corner points of the convex coverage set (CCS)~\cite{roijers2015computing}, named PPO\_CCS. All these results are based on the MuJoCo simulator~\cite{todorov2012mujoco}.

In the first scenario, our goal is to make a three-dimensional bipedal robot walk forward as fast as possible while saving cost simultaneously in the Humanoid-v2 environment. More specifically, our goal is to maximize the Mean Episode Length (MEL), the Mean Episode Reward (MER), and the Linear Velocity (LVel), while minimizing the Quadratic Control (QCtrl) and the Quadratic Impact Cost (QIm). In the second test scenario, our goal is to make a three-dimensional bipedal robot stand up as fast as possible while saving cost simultaneously in the HumanoidStandup-v2 environment. More specifically, our goal is to maximize the standup cost while minimizing the quadratic control and the quadratic impact cost.

\begin{figure*}[htbp]
\centering
\subfigure[Humanoid-v2]{\includegraphics[width=7cm]{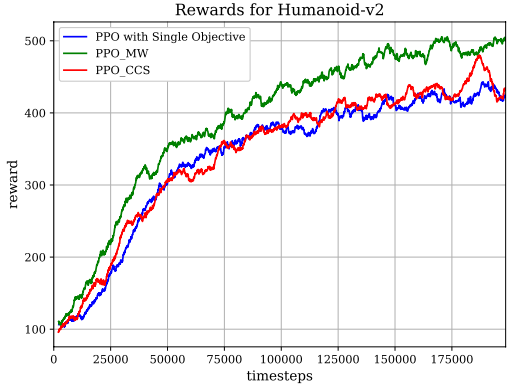}\label{5a}}
\hspace{.2in}
\subfigure[HumanoidStandup-v2 ]{\includegraphics[width=7cm]{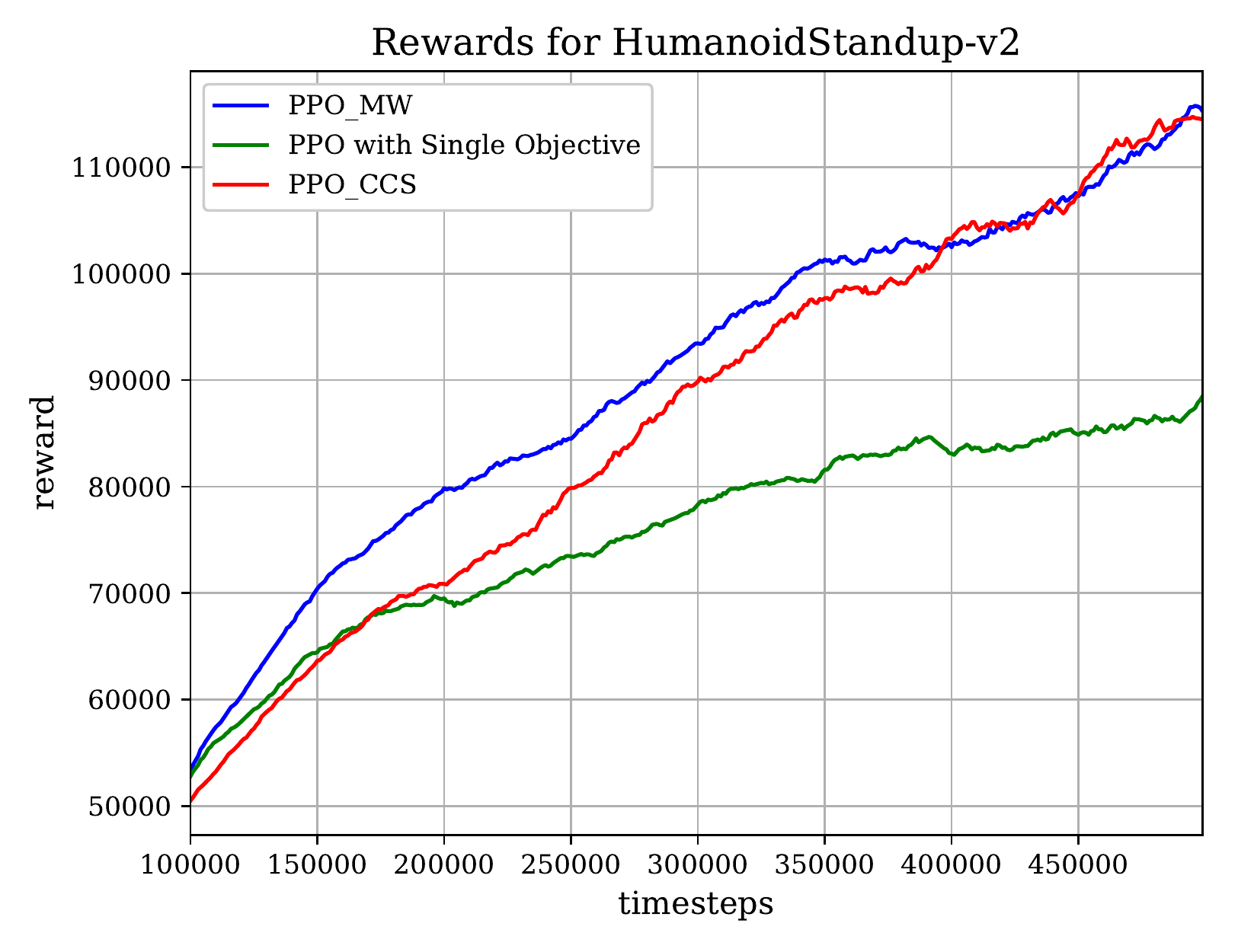}\label{5b}}
%
%
%
\caption{Comparison among PPO\_MW, PPO\_CCS, and PPO with single objective on two test scenarios.}
\label{figure5:p}
\label{Training}
\end{figure*}

We take the current reward function in the OpenAI Gym environments as a baseline, use the cumulative reward trained by the single objective PPO as a benchmark \cite{brockman2016openai,dhariwal2017openai}, and compare it with our proposed method. It is worthwhile to emphasize that HumanoidStandup-v2 does not have a specified reward threshold beyond which ``stand up'' is considered successful. The training results of the experiments are shown in Figure \ref{Training}.

\begin{figure*}[htbp]
\centering
\includegraphics[width=7.2cm]{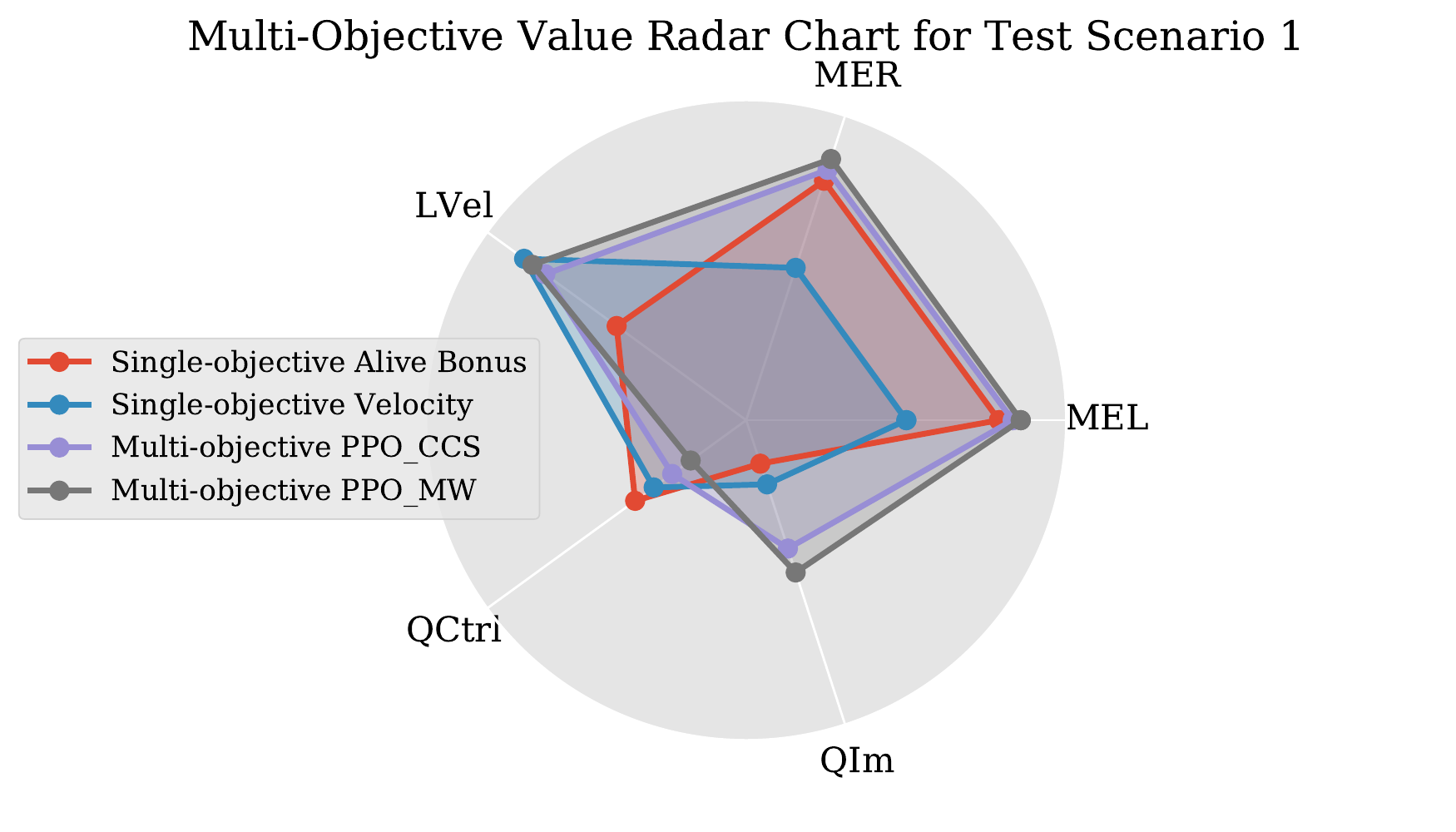}
\includegraphics[width=7.2cm]{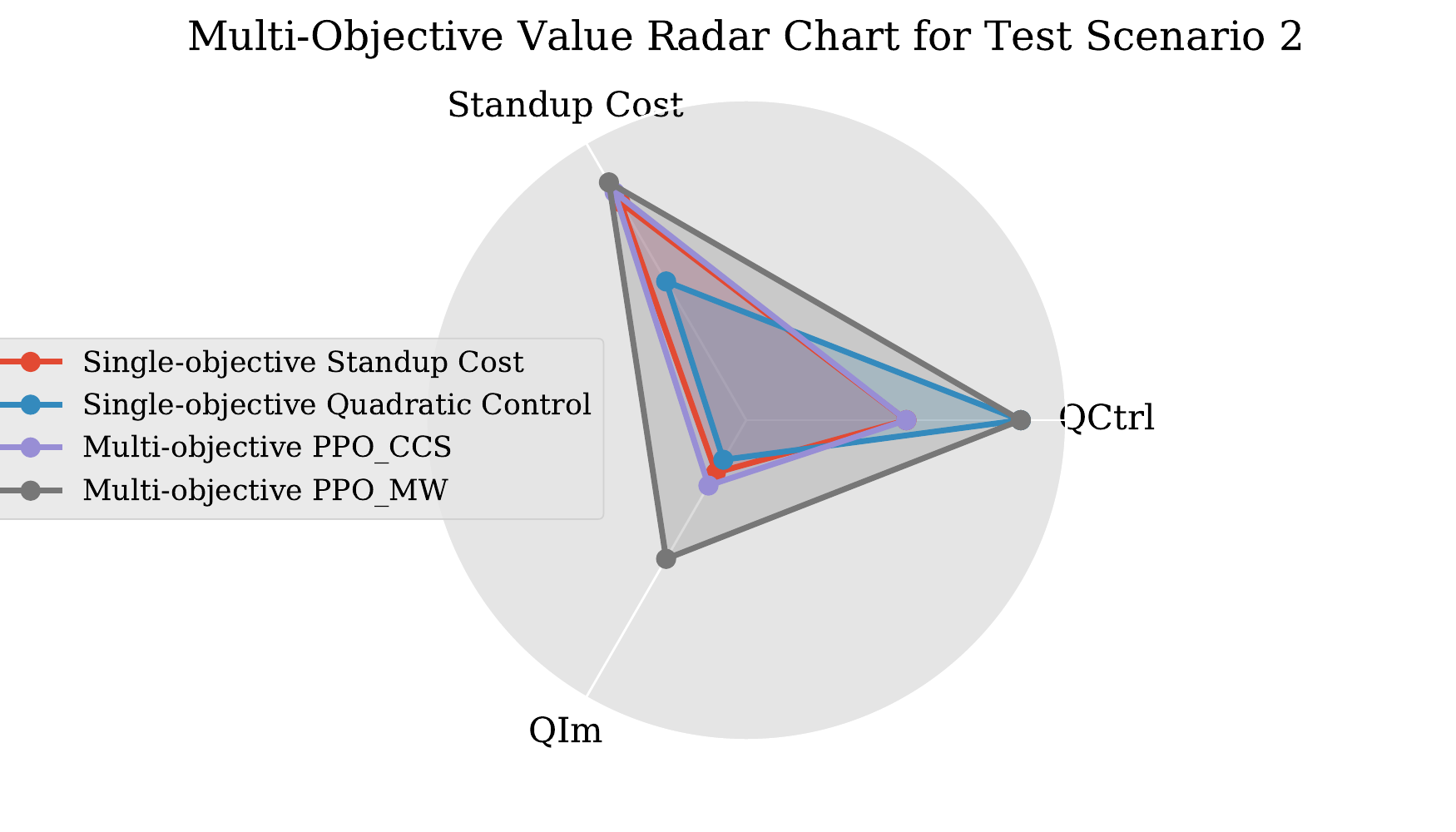}
\caption{A descriptive radar chart for multi-objective values based on several methods.}
\label{radar}
\end{figure*}

For the first scenario, Fig.~\ref{5a} shows the rewards using PPO\_MW, PPO\_CCS, and PPO with single objective. It can be observed that the proposed PPO\_MW yields the highest reward while both PPO\_CCS and PPO with single objective yield lower rewards. Similarly, the rewards using PPO\_MW, PPO\_CCS, and PPO with single objective for the second scenario shown in Fig.~\ref{figure5:p} demonstrate that PPO\_MW outperforms both PPO\_CCS and PPO with single objective. We then test the generalized performance in both scenarios using the final learned policy. The results are shown in Table~\ref{mot1} and Table~\ref{mot2}. The radar charts for illustrative performance comparison are shown in Fig.~\ref {radar}. It can be observed from Table~\ref{mot1} and Table~\ref{mot2} that PPO\_MW can generate higher rewards in most cases because PPO\_MW can optimize multiple objectives simultaneously, as also shown in the radar charts, while the PPO with single objective does not seek to optimize multiple objectives. Meanwhile, because MW can provide more ``weights'' to select than CCS, PPO\_MW can outperform PPO\_CCS even if PPO\_CCS also uses vector value functions.

\begin{table*}
\centering
\setlength{\tabcolsep}{1pc}
\newlength{\digitwidth} \settowidth{\digitwidth}{\rm 0}
\catcode`?=\active \def?{\kern\digitwidth}
\caption{Multi-objective Value for Test Scenario 1}
\label{mot1}
\label{tab:effluents}
\scalebox{0.8}{
\hspace{-2cm}
\begin{tabular*}{\textwidth}{@{}l@{\extracolsep{\fill}}rrrr}
\cline{1-5}
                 & \multicolumn{2}{l}{Single-objective}
                 & \multicolumn{2}{l}{Multi-objective} \\
\cline{2-3} \cline{4-5}
                 & \multicolumn{1}{r}{Alive Bonus}
                 & \multicolumn{1}{r}{Velocity}
                 & \multicolumn{1}{r}{CCS}
                 & \multicolumn{1}{r}{MW}         \\
\cline{1-5}
Mean Episode Length    & $ 60.162 \pm 15.073 $ & $47.670\pm11.943$ & $  62.013\pm15.581$ & $  \mathbf{63.384\pm15.966}$ \\
Mean Episode Reward & $ 405.643 \pm 87.661$ & $45.53 \pm 11.383$ &  $463.186\pm91.253$   & $  \mathbf{506.832\pm92.401}$ \\
Quadratic Control      & $\mathbf{-0.235\pm0.070}$ & $-0.232\pm0.012$ & $ -0.231\pm0.059$ & $ -0.23\pm0.041$ \\
Linear Velocity          & $ 0.310\pm0.262$ & $\mathbf{1.013\pm0.481}$ &   $0.385\pm0.253$      & $  0.490\pm0.197$ \\
Quadratic Impact Cost           & $ -0.43\pm0.14$ & $-0.48\pm0.09$ & $  -0.64\pm0.04$ & $  \mathbf{-0.70\pm0.05}$ \\
\cline{1-5}
\multicolumn{5}{@{}p{160mm}}{Average and standard deviation $\left(\mu\pm\sigma\right)$ of multi-objective values, the best testing performance is shown in boldface. Radar chart of multi-objective value per attribute for Humanoid-v2 is shown in Fig.~\ref{radar}. }
\end{tabular*}}
\end{table*}

\begin{table*}
\centering
\setlength{\tabcolsep}{1pc}
\catcode`?=\active \def?{\kern\digitwidth}
\caption{Multi-objective Value for Test Scenario 2}
\label{mot2}
\label{tab:effluents}
\scalebox{0.7}{
\hspace{-4cm}
\begin{tabular*}{\textwidth}{@{}l@{\extracolsep{\fill}}rrrr}
\cline{1-5}
                 & \multicolumn{2}{l}{Single-objective}
                 & \multicolumn{2}{l}{Multi-objective} \\
\cline{2-3} \cline{4-5}
                 & \multicolumn{1}{r}{Standup Cost}
                 & \multicolumn{1}{r}{Quadratic Control}
                 & \multicolumn{1}{r}{CCS}
                 & \multicolumn{1}{r}{MW}         \\
\cline{1-5}
Quadratic Control      & $-0.215\pm0.027$ & $\mathbf{-0.216\pm0.039}$ & $ -0.215\pm0.026$ & $\mathbf{ -0.216\pm0.027}$ \\
Standup Cost          & $ 74417.100\pm16135.020$ & $789.094\pm212.467$ &   $85793.17\pm18951.362$      & $ \mathbf{88899.880\pm18934.457}$ \\
Quadratic Impact Cost           & $ -0.210\pm0.012$ & $-0.207\pm0.024$ & $  -0.227\pm0.016$ & $  \mathbf{-0.230\pm0.015}$ \\
\cline{1-5}
\multicolumn{5}{@{}p{170mm}}{Average and standard deviation $\left(\mu\pm\sigma\right)$ of multi-objective values, the best testing performance is shown in boldface. Radar chart of multi-objective value per attribute for HumanoidStandup-v2 is shown in Fig.~\ref{radar}.}
\end{tabular*}
}
\end{table*}

\subsection{Solution Stability}\label{Stability}
\begin{figure*}[htbp]
\centering
\subfigure[Episode 89]{\includegraphics[width=5cm]{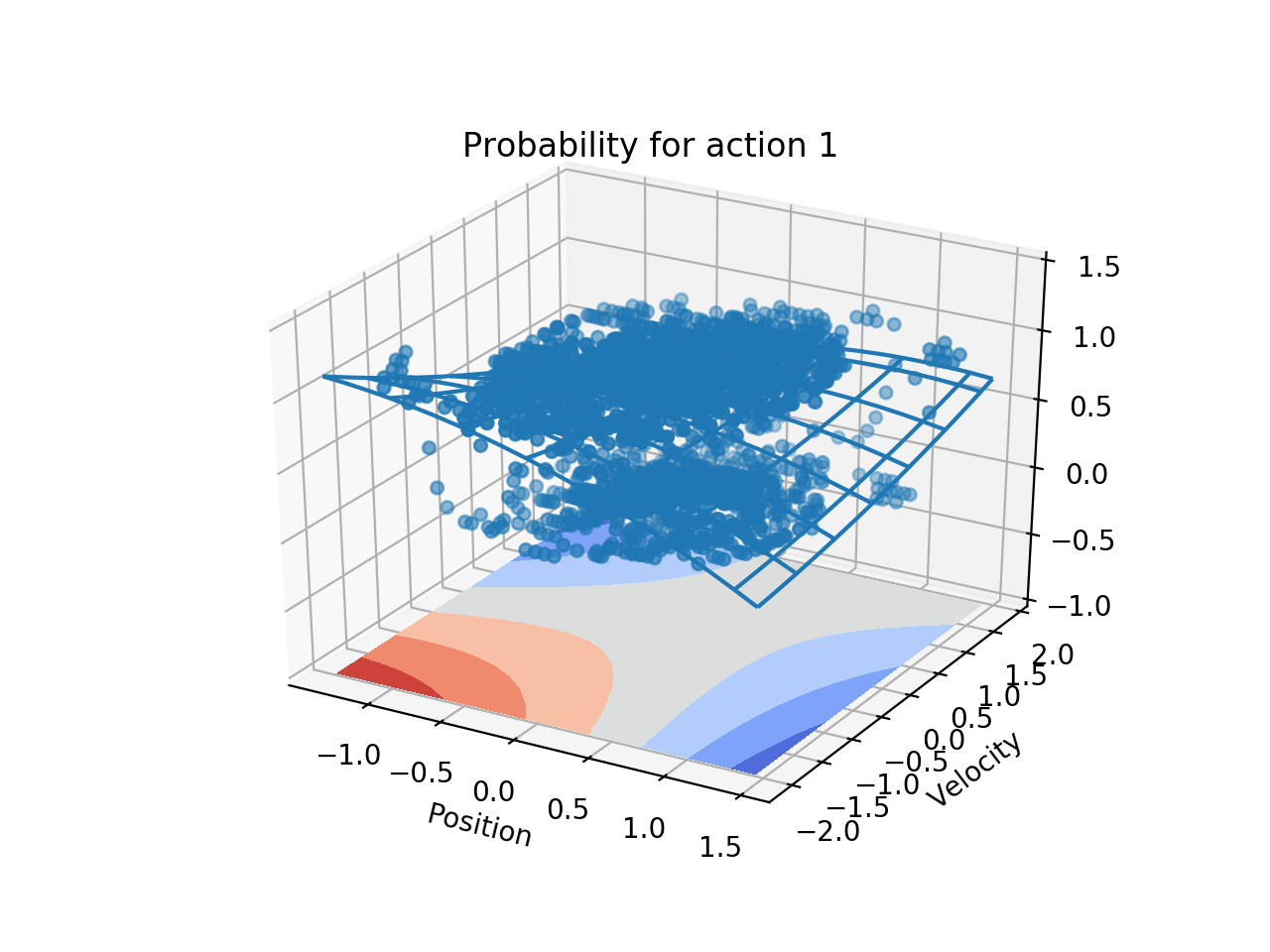}}
\subfigure[Episode 120]{\includegraphics[width=5cm]{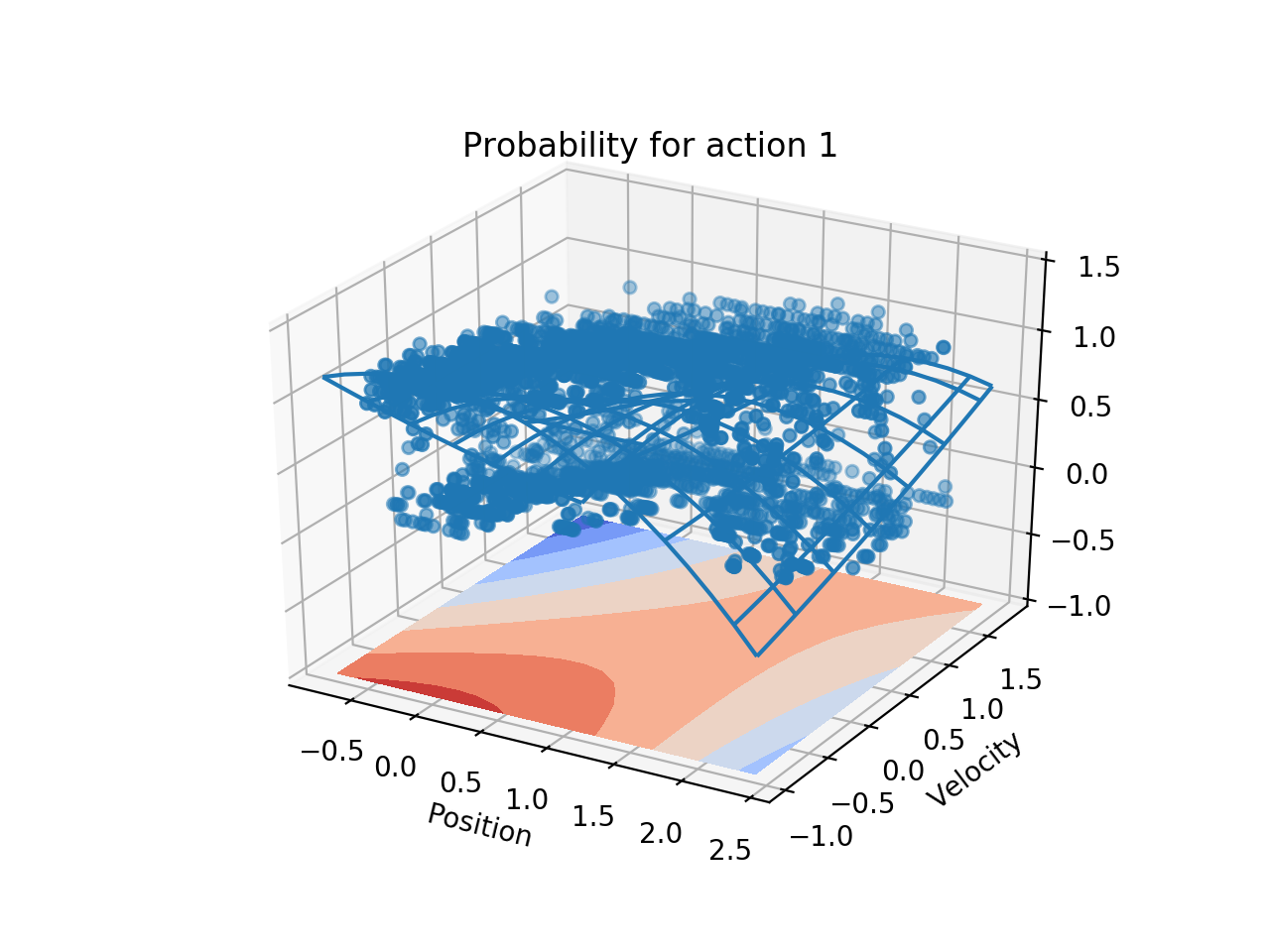}}
\subfigure[Episode 157]{\includegraphics[width=5cm]{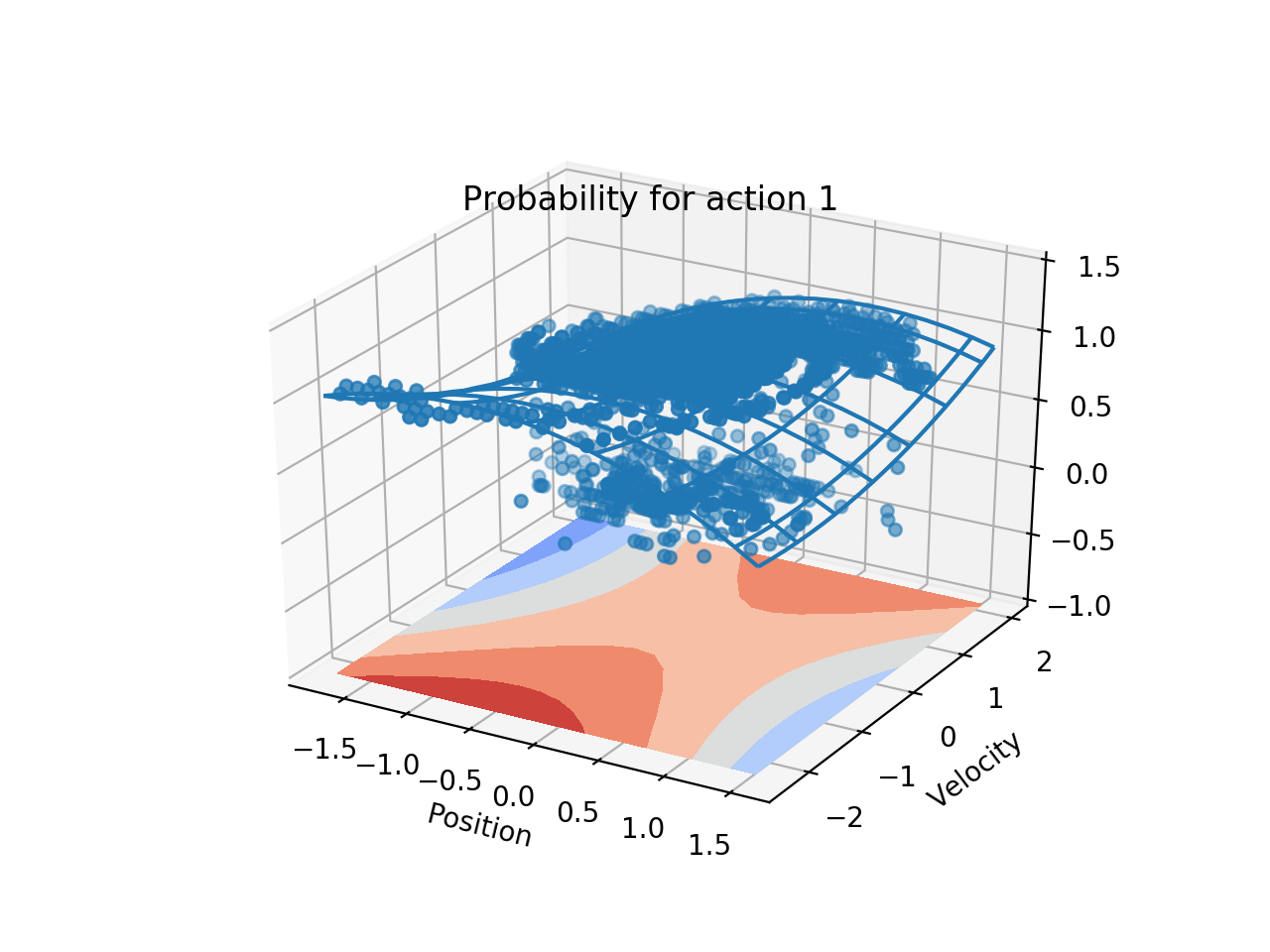}}
\caption{The actor $\pi\left ( s;\theta_{\pi} \right )$ trained on the Humanoid-v2. The surfaces represent the functions. The blue dots show the trajectories in the state-action using the current policy. The red and blue contours show the forces that shape the surfaces.}
\label{pc}
\end{figure*}

We finally show that the proposed method can provide stable solutions. Fig. \ref{pc} shows the learned policy at the $89$th, $120$th, and $157$th episodes for the first scenario. The red and blue contours show the forces, $\sum_{k}\bigtriangledown _{\theta _{\pi}}\log \pi_{\theta}\left ( s_{k},a_{k} \right ) \delta_{k,t}$, that shape the surfaces. The lines show the probabilities for one particular action. After the $157$th episode, the contours show that the learned policy become more and more stable because the policy becomes more consistent across episodes, indicating that the policy is ``settled'' after an appropriate number of episodes. 

It is worth noting that the stability of the proposed method is determined by both the stability of the actor-critic network and the stability in learning $W$. Moreover, the stability of the actor-critic network and the stability in learning $W$ are interdependent. On the one hand, if the actor-critic network is stable, $W$ can often be stabilized since $W$ becomes the main one to be learned after the stabilization of the actor-critic network. On the other hand, if $W$ is stable, the actor-critic network becomes the main one to be trained after the value of $W$ is stabilized. Notice that $W$ described in Subsection~\ref{CM} becomes stable as the training proceeds. Hence, the actor-critic network will also become stable, which is consistent with the obtained stable action policy shown in Fig.~\ref{pc}. 


\section{Discussion and Conclusion}
In multi-objective optimization problems, the possibly conflicting objectives necessitates a trade-off when multiple objectives need to optimize simultaneously. A typical approach is to minimize a loss of weighted linear summation of all objective functions. However, this approach may be effective for limited cases (e.g., when the objectives do not compete). To address the potential completing nature among these objectives, we proposed a vector value function based multi-objective deep reinforcement learning to solve high-dimensional multi-objective decision making problems in continuous control environments. The proposed method optimizes vectorized proxy objectives sequentially based on proximal policy optimization, actor-critical network, and the derivation of optimal weights via marginal weight. 

By explicitly quantifying inter-objective relationship via correlation matrix, the relative importance of the objectives unknown \textit{a prior} can be obtained via reinforcement learning. Each entry in the correlation matrix specifies and explains the relative impact of one objective on another objective in the optimization step. After presenting the main approach, we also proved its convergence and demonstrated its advantages via two testing scenarios in the MuJoCo environment. 

There remain many interesting directions for future research into multi-objective optimization. For example, how to evaluate the impact of a perturbation on a row of $W$ on individual objective $V_{i}^{\pi}$? How to consider constraints on selected objectives? 


\bibliography{references1} 
\bibliographystyle{ieeetr}

\appendix

\section*{A.1 The Pseudocode for AOLS}\label{sec:AOLS}

\begin{algorithm}[htbp]
\KwData{MOMDP: m, improvement threshold: $\varepsilon$}
\KwResult{US, $\Delta_{max}$}
\emph{S: empty partial US, W: empty list of explored marginal weight, Q: an empty priority queue of the initial marginal weight, $\Delta_{max}$: improvement}\\
\ForAll{extreme weights of infinite priority $w_{\max} =e_{1}$}
 {Q.add $(w_{\max},\infty)$}
\While{$\neg$ Q.isEmpty()  $\wedge \neg$ timeOut}{
$w_{i}^{j} \leftarrow$Q.pop()\\
$WV_{old} = WV_{old}  \cup  \left \{ \left ( w_{i}^{j} , w_{i}^{j}\cdot V(s,\phi_V) \right)\right \}$\\
\If{$ V(s,\phi_V) \notin S $}{
$S\leftarrow S \cup \left \{ V(s,\phi_V)  \right \}$\\
$W\leftarrow$ recompute marginal weight $V_{S}^{*}\left ( w \right )$\\
\For{$K\in 1,...,len(w)$}{
\If{$e_{K}\neq W$}{return$\left(e_{K},\infty \right)$}}
$V_{US}\left [ \cdot  \right ]\leftarrow$ $\forall$ weights in $W\left [ \cdot \right ]$, compute: max $w_{i}^{j}[K]\cdot v(s,\phi_V)$\\
subject to: $\forall \left (w_{i}^{j}[K],u \right )\in WV: w_{i}^{j}[K]\cdot v(s,\phi_V)\leq u+\varepsilon$\\$K\leftarrow \arg\max_{K}  V_{US}\left [ K \right ] - V_{S}^{*}\left ( W\left[K\right] \right )$\\
\If{$ V_{US}\left [ K \right ] - V_{S}^{*}\left ( W\left[K\right] \right ) > \varepsilon $}
{ Q.add$(W[K], V_{US}\left [ K \right ] - V_{S}^{*}\left ( W\left[K\right] \right ))$}}
$W\leftarrow W\cup \left \{ W\left [ K \right ] \right \}$
}
\caption{function AOLS$\left( m, \varepsilon,  V(s,\phi_k)\right)$}
\label{alg:DOL}
\end{algorithm}

\clearpage
\section*{A.2 The Pseudocode for the Proposed Algorithm} \label{sec:alg}

\begin{algorithm}[htbp]
\KwData{initial policy parameters $\theta_{0}$, initial value function parameters $\phi_{0}^{V_{i}}$, initial vectorized weights based on each objective $w_{ij}$, $US$}
\KwResult{PPO\_Model}
\For{$i=1$ to $I$}{
\For{$k=0,1,2,...,K$}
{ Collect set of trajectories $D_{k}^{i}=\left \{ \tau _{n}^{i} \right \}$ by running policy $\pi_{k}=\pi\left ( \theta_{k} \right )$ in the environment\\
Compute rewards-to-go $\hat{R_{t}}$\\
Update $ V(s,\phi_k)$\\
 Compute $w_{i\left [ \cdot \right ] } $ by function AOLS$\left( m, \varepsilon, V(s,\phi_k) \right)$\\
 $V_{i}^{\phi_{k}} = w_{i\left [ \cdot \right ] } \times  V(s,\phi_k) $\\
Compute advantage estimates $\hat{A_{t}^{i}}$ using GAE method based on the current value function $V_{i}^{\phi_{k}}$\\
 Update the policy by maximizing the PPO-Clip objective:
\begin{align*}
\theta_{k+1} = &\underset{\theta}{\arg\max}\frac{1}{\left |D _{k}^{i} \right | T}\sum_{\tau_{n}^{i}\in D _{k}^{i} } \sum_{t=0}^{T}\min\Big( \frac{\pi_{\theta}\left ( a_{t}|s_{t} \right )}{\pi_{\theta_{k}}\left ( a_{t}|s_{t} \right )} \\
& A^{i,\pi_{\theta_{k}}}\left( s_{t},a_{t} \right ), g\left ( \epsilon, A^{i,\pi_{\theta_{k}}}\left( s_{t},a_{t} \right )  \right ) \Big)
\end{align*}
via stochastic gradient ascent (e.g., Adam)\\
Fit value function by regression on mean-squared error:
\begin{align*}
\phi_{k+1}^{V_{i}}=\underset{\phi_{k}^{V_{i}}}{\arg\min}\frac{1}{\left | D_{k}^{i} \right |T}\sum\limits_{\tau_{n}^{i}\in D _{k}^{i} } \sum\limits_{t=0}^{T}\left ( V_{i}^{\phi_{k}} \left ( s_{t} \right )-\hat{R_{t}^{i}} \right )^{2}
\end{align*} via stochastic gradient descent.}
}
\caption{Pseudocode for the calculation of vectorized multi-objective values}
\label{alg:seq}
\end{algorithm}

\section*{A.3 Proof of Theorem~\ref{th:main}} \label{sec:conv}

We first start with some notations used in the proof. Consider a Markov decision process with finite state space $X$, and finite action space $\mathcal{A}$. Let $R:X\times \mathcal{A}\rightarrow\mathbb{R}$ be a given reward function and $\mu$ be a mapping from each observation $x$ to the action space $\mathcal{A}$ as a probability distribution. In particular, we use $\mu_a(x,\varphi),~\varphi \in \mathbb{R}^{n},$ to represent the probability of selecting action $a$ at state $x$ subject to the parameter $\varphi$ that corresponds to a policy $\pi_x(\varphi)$. We define one stage reward as $g_{x_k}(\varphi)$, where $x_k$ is the state at the time step $k$ (one-stage state). The performance metric that we used to compare different policies is the average reward criterion given by 
$$\lambda(\varphi)=\underset{t\rightarrow\infty}{\lim}\frac{1}{t}E\left[\sum^{t}_{k=0}g_{x_k}(\varphi)\right],$$
where $E$ is the expectation operator. 
For any $\varphi $ and $x$, the differential reward $v_x(\varphi)$ of observation $x$ is defined as 
$$v_x(\varphi)=E\left[\sum_{k=1}^{T-1}\left(g_{x_k}(\varphi)-\lambda(\varphi)\right)|x_0=x\right],$$ 
where $T=\min\left \{ k>0|x_k\in \{x_i,i=0,\cdots,k-1\}\right \}$.

Before deriving an explicit form of the gradient of $\lambda(\varphi)$, we make the following assumptions:
\begin{assumption}\label{ape}
For each $\varphi$, the Markov chains $\left\{X_n\right\}$ and $\left\{X_n,\mathcal{A}_n\right\}$, denoting the sequence of states and state-action pairs, are irreducible and aperiodic under the stationary probabilities $\pi_x(\varphi)$ and $\eta_a(x,\varphi)= \pi_x(\varphi)\mu_a(x,\varphi)$.
\end{assumption}
\begin{assumption}\label{bou}
For every $x,x^\prime \in X$, the transition probability $p_{xx^\prime}(\varphi)$ and $g_{x}(\varphi)$ are bounded, twice differentiable, and have bounded first and second derivatives. In addition, for every observation $x$ and action $a$, there exists a bounded function such that
\begin{equation}
\psi_a(x,\varphi) = \frac{\bigtriangledown\mu_a(x,\varphi)}{\mu_a(x,\varphi)}, 
\end{equation}
where $\psi_a(x,\varphi)$ has first bounded derivatives for any fixed $x$ and $a$.
\end{assumption}


\begin{lemma}\label{lem:gradient}
Let Assumptions \ref{ape} and \ref{bou} hold true. Then the gradient of $\lambda(\varphi)$ can be represented as
\begin{equation}\label{innerProduct}
\bigtriangledown\lambda(\varphi_i)=\sum_{x\in X}\sum_{a\in \mathcal{A}}\eta_a(x,\varphi)q_{x,a}(\varphi)\psi^i_a(x,\varphi),
\end{equation}
where
$
q_{x,a}(\varphi)=(g_{x,a}-\lambda(\varphi))+\sum_{x^\prime \in X} p_{xx^\prime}(a)v_{x^\prime}(\varphi)
$
and $\psi^i_a(x,\varphi)$ is the $i$th component of $\psi_a(x,\varphi)$.
\end{lemma}
\proof Note that the gradient of $\lambda(\varphi)$ is given by~\cite{marbach2001simulation} 
\begin{equation}\label{lamdaVarphi}
\bigtriangledown\lambda(\varphi)=\sum_{x \in X}\pi_x(\varphi)\left(\bigtriangledown g_{x}(\varphi)+\sum_{x^\prime \in X}\bigtriangledown p_{xx^\prime}(\varphi)v_{x^\prime}(\varphi)\right).
\end{equation}
The expected reward per stage $g_{x}(\varphi)$ is given by
\begin{equation*}
g_{x}(\varphi)=\sum_{a \in \mathcal{A}}\mu_a(x,\varphi)g_{x,a}.
\end{equation*} 
Then the gradient of $g_{x}(\varphi)$ can be written as
\begin{align*}
\bigtriangledown g_{x}(\varphi)&=\sum_{a \in \mathcal{A}}\bigtriangledown\mu_a(x,\varphi)g_{x,a}\\
&=\sum_{a \in \mathcal{A}}\bigtriangledown\mu_a(x,\varphi)g_{x,a}- \bigtriangledown\sum_{a \in \mathcal{A}}\mu_a(x,\varphi)\lambda(\varphi)
\end{align*}
because $\sum_{a \in \mathcal{A}}\mu_a(x,\varphi)=1$ and hence $\bigtriangledown\sum_{a \in \mathcal{A}}\mu_a(x,\varphi)=0$. Then we can further obtain that
\begin{align}
\bigtriangledown g_{x}(\varphi)&=\sum_{a \in \mathcal{A}}\bigtriangledown\mu_a(x,\varphi)g_{x,a}- \sum_{a \in \mathcal{A}}\bigtriangledown\mu_a(x,\varphi)\lambda(\varphi) \notag\\
&=\sum_{a \in \mathcal{A}}\bigtriangledown\mu_a(x,\varphi)(g_{x,a}-\lambda(\varphi))\label{eq:grad_g}
\end{align}
by moving the gradient inside the summation.
Meanwhile, the transition probability is given by: 
\begin{equation}
p_{xx^\prime}(\varphi)=\sum_{a \in \mathcal{A}}\mu_a(x,\varphi) p_{xx^\prime}(a).
\end{equation}
By following a similar analysis as that for $\bigtriangledown g_{x}(\varphi)$, we can obtain:
\begin{equation} \label{eq:grad_p}
\sum_{x^\prime \in X}\bigtriangledown p_{xx^\prime}(\varphi)v_{x^\prime}(\varphi)=\sum_{x^\prime \in X}\sum_{a \in \mathcal{A}}\bigtriangledown\mu_a(x,\varphi) p_{xx^\prime}(a)v_{x^\prime}(\varphi).
\end{equation}
By inserting~\eqref{eq:grad_g} and~\eqref{eq:grad_p} int \eqref{lamdaVarphi} and making a few rearrangements, we can obtain~\eqref{innerProduct}. 
\endproof

Based on Lemma~\ref{lem:gradient}, we now show that the gradient $\bigtriangledown\lambda(\varphi)$ can be written in the form of inner products given in the following lemma. Before moving on, let's define $q_\varphi$ and $\psi(\varphi)$ as the vectors of, respectively, $q_{x,a}(\varphi)$ and $\psi_a(x,\varphi)$ on $X\times \mathcal{A}$. Define the inner product of two real value functions $q_\varphi$ and $\psi(\varphi)$ as
\begin{equation}\label{eq:innerproduct}
\left \langle q_\varphi,\psi(\varphi) \right \rangle_\varphi=\sum_{x\in X}\sum_{a\in \mathcal{A}}\eta_a(x,\varphi)q_{x,a}(\varphi)\psi_a(x,\varphi).
\end{equation}

\begin{lemma}\label{lem:equiv}
The gradient of $\lambda(\varphi)$ can be computed by the inner product of two real value functions given by
\begin{equation}\label{eq:innp}
\bigtriangledown\lambda(\varphi)=\left \langle q_\varphi,\psi(\varphi) \right \rangle_\varphi= \left \langle \prod_{\varphi}q_\varphi,\psi(\varphi) \right \rangle_\varphi,
\end{equation}
where 
\begin{equation}\label{eq:prj}
\prod_\varphi q=\arg\underset{\widehat{q} \in \zeta_\varphi}{\min}\left \|q-\widehat{q}  \right \|_\varphi
\end{equation} 
with $\zeta_{\varphi}$ denoting the span of the vectors $\left \{ \psi^i_a(x,\varphi); i=1,\cdots,n \right \}$ in $\mathbb{R}^{X\times \mathcal{A}}$.
\end{lemma}
\proof Based on the definition in~\eqref{eq:innerproduct} and Lemma~\ref{lem:gradient}, we can obtain that $\bigtriangledown\lambda(\varphi)=\left \langle q_\varphi,\psi(\varphi) \right \rangle_\varphi$.
We next show that the second equality in~\eqref{eq:innp} holds. 

We can rewrite \eqref{innerProduct} by:
\begin{equation}\label{eq:partip}
\frac{\partial }{\partial \varphi_i}\lambda(\varphi)=\left \langle q(\varphi),\psi^i(\varphi) \right \rangle_\varphi,~~~i=1,\cdots,n,
\end{equation}
where $n$ is the dimension of $\varphi$. For a high dimensional space, computing the gradient of $\lambda(\varphi)$ depends on $q_{x,a}(\varphi)$, (equivalently $q_\varphi$ in~\eqref{eq:partip}), and is typically difficult. An alternative approach is to use the project of $q_\varphi$ based on~\eqref{eq:prj} in the computation of inner product. Based on~\eqref{eq:innerproduct}, the inner product $\left \langle q_\varphi,\psi(\varphi) \right \rangle_\varphi$ is equivalent to the inner product of $\psi(\varphi)$ and the projection of $q_\varphi$ on $\zeta_{\varphi}$. Hence, $\left \langle q_\varphi,\psi(\varphi) \right \rangle_\varphi= \left \langle \prod_{\varphi}q_\varphi,\psi(\varphi) \right \rangle_\varphi$ always holds.
In other words, the projection of $q_\varphi$ onto $\zeta_\varphi$ is sufficient to learn $\bigtriangledown\lambda(\varphi)$ since $\left \langle q_\varphi,\psi(\varphi) \right \rangle_\varphi= \left \langle \prod_{\varphi}q_\varphi,\psi(\varphi) \right \rangle_\varphi$.
\endproof

With Lemma~\ref{lem:equiv}, we are ready to present the proof of Theorem~\ref{th:main}. Before moving on, the following two assumptions are needed.
\begin{assumption}\label{ass:stepsize}
The value update stepsize sequence for the critic $\left \{ \gamma^i_k\right \}$  and the actor $\left \{ \beta_k\right \}$ are positive and nonincreasing, and satisfies $\sum_{k=0}^\infty\vartheta_k=\infty$ and $\sum_{k=0}^\infty\vartheta^2_k<\infty$, $\vartheta\in\{\gamma,\beta\}$. In addition, the actor updates much slower that the critic, \textit{i.e.,} $\frac{\beta_k}{\gamma^i_k}\rightarrow 0$.
\end{assumption}

\begin{assumption}\label{ass:Phi}
For every $\varphi_k \in \mathbb{R}^n$, $\Phi \varphi_k \neq e$, where $e$ is equal to all-one vector and $\Phi$ is a $m\times n$ matrix whose $k$th row is equal to $\varphi_k$. In addition, the column vectors of $\Phi$ are linearly independent.
\end{assumption}

\underline{\textbf{Proof of Theorem~\ref{th:main}:}} Under Assumption~\ref{ass:stepsize}, the size of actor updates is negligible compared with the size of the critic updates. If the critic network is stable, the actor network is stationary. We next show the convergence of critic network.

As shown in Lemma~\ref{lem:equiv}, the gradient $\bigtriangledown\lambda(\varphi)$ can be written in the form of inner products. When Assumptions~\ref{ass:stepsize} and~\ref{ass:Phi} hold, the convergence analysis in \cite{tsitsiklis1999average} can be used to prove that any critic in the proposed policy will converge with probability $1$. Once the critic networks converge (\textit{i.e.,} are stationary), all hyperparameters in the critic networks are stationary. In this case, the update of the actor network can be proved stationary as well by the stochastic approximation algorithm~\cite{spall1992multivariate}. 

This completes the proof of Theorem~\ref{th:main}.
\endproof

\end{document}